\documentclass[useAMS,usenatbib    
]{mn2e}    
\usepackage{times}    
\usepackage{graphicx}    
\usepackage{amsmath}    
\usepackage{natbib}    
\usepackage{color,ulem}  
\usepackage{amssymb}         
\bibpunct{(}{)}{;}{a}{}{,}    
         
\title[Ultra-Compact Dwarfs around NGC\,3268]{Ultra-Compact Dwarfs around NGC\,3268 \thanks{Based on  
observations collected at the Cerro   
Tololo Interamerican Observatory (CTIO); observations obtained at the Gemini    
Observatory, which is operated by the Association of Universities for Research in    
Astronomy, Inc., under a cooperative agreement with the NSF on behalf of the Gemini    
partnership: the National Science Foundation (United States), the Science and    
Technology Facilities Council (United Kingdom), the National Research Council (Canada),    
CONICYT (Chile), the Australian Research Council (Australia), Minist\'erio da Ciencia e    
Tecnologia (Brazil) and Ministerio de Ciencia, Tecnolog\'ia e Innovaci\'on Productiva    
(Argentina); and observations carried out at the European Southern Observatory, Paranal    
(Chile), programme 71.B-0122(A).}}   
   
\author[Juan Pablo Caso et al.]     
{Juan Pablo Caso$^{~1,2}$\thanks{E-mails:\,jpcaso@fcaglp.unlp.edu.ar\,(JPCas);    
\,lbassino@fcaglp.unlp.edu.ar\,(LPB);\,tom@astro-udec.cl\,(TR);    
\,jpcalderon@fcaglp.unlp.edu.ar (JPCal;\,asmith@fcaglp.unlp.edu.ar\,(ASC);}, Lilia P. Bassino$^{~1,2}$, Tom Richtler$^{~3}$,  
\newauthor 
Juan Pablo Calder\'on$^{~1,2}$ and Anal\'ia V. Smith Castelli$^{~1,2}$\\    
$^{1}$Grupo de Investigaci\'on CGGE, Facultad de Ciencias Astron\'omicas y Geof\'isicas, Universidad Nacional de La Plata,     
and \\ Instituto de Astrof\'isica de La Plata (CCT La Plata -- CONICET, UNLP), Paseo del Bosque S/N,   
B1900FWA La Plata, Argentina\\    
$^{2}$Consejo Nacional de Investigaciones Cient\'ificas y T\'ecnicas, Rivadavia 1917, C1033AAJ   
Ciudad Aut\'onoma de Buenos Aires, Argentina\\  
$^{3}$Departamento de Astronom\'ia, Universidad de Concepci\'on, Casilla 160--C, Concepci\'on, Chile}   
\begin{document}    
    
\date{Accepted . Received ; in original form }    
    
\pagerange{\pageref{firstpage}--\pageref{lastpage}} \pubyear{2011}    
    
\maketitle    
    
\label{firstpage}    
    
\begin{abstract}   
We present radial velocities (from Gemini/GMOS) of the second sample of 
ultra-compact dwarfs (UCDs) and bright globular clusters (GCs) in the 
Antlia cluster. Twenty-three objects are located around the giant 
elliptical NGC\,3268, and one is close to the fainter lenticular NGC\,3273.
Together with previously found UCDs around NGC\,3258
a total of 35 UCDs and bright GCs has been now identified in the Antlia 
cluster. Their colours and magnitudes are compared with those of the nuclei 
of dE,N galaxies already confirmed as Antlia members. 
For a subsample that lie on ACS images and are brighter than $M_V = -9$\,mag, 
the effective radii ($R_{\rm eff}$) have been measured, the maximum radius being 
approximately 10\,pc.
In addition to the radial velocity sample, we find 10 objects in the magnitude 
range corresponding to GCs but with $10 < R_{\rm eff} < 17$\,pc,
resembling  the so-called `extended clusters'.
By number and magnitude, the new UCDs fit to the GC luminosity function, supporting 
their interpretation as bright GCs.
Additionally, we use a tracer mass estimator to calculate the mass enclosed up to
$\approx 47$\,kpc from NGC\,3268, which results in $2.7\times 10^{12}\,{\rm M_\odot}$.

\end{abstract}    
    
    
\section{Introduction}    
 
The discovery of objects in the Fornax cluster, which were much brighter than ''normal'' 
globular clusters (GCs), but much smaller than dwarf ellipticals \citep{min98,hil99,dri00} 
made a strong impression on the star-cluster community.

The designation Ultra-Compact Dwarf (UCD) has been  introduced by \citet{dri02}
 to emphasize the new nature of these bright and, at least in comparison to dwarf galaxies,
compact objects (see \citealt{hil09b} for a review). When \citet{has05} found strikingly high 
M/L-values for UCDs in the Virgo cluster, the possibility of a dark matter component, which 
would mean a galaxy nature,  was seriously considered. However, further studies found lower 
M/L-values consistent with stellar populations, e.g. \citet{fra11}. In addition
\citet{wil12} state that similar results have been obtained by several dynamical analyses
of UCDs. In spite of that we continue
to use the term ''UCD´´ even if these objects are bright star clusters rather than faint
galaxies and even, if there is no clean definition.

Apparently, cluster-like environments \citep{gre09,bro11,chib11,mis11,pen12,cas13a} are not 
needed to produce UCDs, since they have been found also around individual galaxies
\citep{rej07,hau09,mad13}. For example, the number of clusters brighter than $\omega$ Cen
in NGC 4636 is about 70 \citep{sch12}.

There is no generally accepted definition for UCDs and different authors assign different
luminosity ranges. For example \citet{hil09b} suggested a $V$ absolute magnitude range of 
$-13.5 < M_V < -11$, while \citet{bro11} extended it towards a fainter limit ($M_V < -9$).
More recently \citet{mie12} proposed that UCDs were systems with $M_V < -10.25$. \citet{bru12}
discriminated as UCDs those objects with effective radii ($R_{\rm eff}$) larger than 10\,pc 
and luminosities in the dwarf galaxies' regime, while fainter objects with similar radii were 
called `extended clusters (ECs)'. The simulations of \citet{pfe13} showed that the 
remnants of stripped nucleated dwarf elliptical galaxies (dE,N) could present effective radii 
($R_{\rm eff}$) of only a few parsecs. In this sense, $\omega$\,Cen has been considered as a 
possible remnant of a Milky Way satellite galaxy \citep[e.g.][]{hil00,bek03b,bok08}, and its 
$R_{\rm eff}$ is $\approx 7.5$\,pc \citep[][2010 Edition]{har96}. This scenario is 
supported on the evidence of multiple stellar populations formed along several Gyrs 
\citep[e.g.][]{lee99,bed04,hil04,mar11}, and the
presence of Galactic tidal streams that can be associated with $\omega$\,Cen \citep{maj12}.
Moreover, \citet{ols09} found that NGC\,1851, a faint Galactic GC with
$R_{\rm eff} \approx 2$\,pc, is surrounded by a diffuse stellar halo. This has been 
pointed out as evidence that the origin of NGC\,1851 may have been the disruption of a dwarf 
galaxy \citep{bek12}. 
Consequently, we may loose UCD candidates if we use $R_{\rm eff}$ as a criterion. We will select as UCDs those systems with similar 
colours than GCs, and $M_V < -10.5$.

The origin of UCDs is not clear either. A substantial fraction probably constitute the bright 
end of the GC population of their host galaxy \citep[e.g.][]{nor11,mie12}, but there may be 
various formation channels \citep[e.g.][]{hil09a,hil09b,chil11,bru12}. Tidal stripping of 
nucleated galaxies has been proposed as a possible origin for some UCDs based on  observational 
studies \citep[e.g.][]{str13}, but also  from numerical simulations \citep{bas94,bek01,bek03a,pfe13}.

Although most UCDs show M/L-values consistent with stellar populations, the discussion about 
higher dynamical M/L-values, particularly of the brighter objects, is still ongoing
 \citep{mie08,tay10,str13}. Rather than dark matter halos, top-heavy initial mass functions (IMF)
\citep{dab09,mur09,dab12}, or central black holes \citep{mie13} have been suggested as
responsible for those high $M/L$-values.

\subsection{The Antlia cluster}    
 
The present work has been performed within the context of the Antlia Cluster Project that aims 
at studying the galaxy content of this cluster, i.e. the GCs associated to the two dominant 
elliptical galaxies \citep{dir03b,bas08}, the complete galaxy population \citep[][Calder\'on et 
al., in prep.; Bassino et al., in prep.]{smi08a,smi08b,smi12}, and the UCDs \citep[][hereafter 
Paper\,I]{cas13a}. 
 
The Antlia galaxy cluster is the third  nearest well populated galaxy cluster. It consists 
mainly of two groups, each one dominated by a giant elliptical (gE) galaxy (NGC\,3258 and 
NGC\,3268). Considered as an example of a galaxy cluster in an intermediate merger stage 
\citep{haw11}, the Antlia cluster is a very interesting target for studying the UCD population.
In Paper\,I we analyzed a sample of confirmed UCDs and marginally resolved candidates around  
NGC\,3258, and we obtained the photometric properties for a sample of Antlia dE,N galaxies.
 It was shown that dE,N nuclei and UCDs occupy the same locus in the $V,I$ colour-magnitude 
diagram (CMD). The projected spatial distribution of UCDs proved to be similar to those of 
NGC\,3258 GCs. In addition, effective radii of UCDs were measured  with ACS (Advanced Camera 
for Surveys) imaging available from the {\it Hubble Space Telescope} (HST) archive, giving 
support to the existence of a size-luminosity relation.  
\medskip   
 
The structure of this paper is the following. Section\,2 describes the observations,    
reductions, and the adopted criteria for the selection of GCs and UCDs. In Section\,3    
we present the results regarding their colour-magnitude relation (CMR),  
size, size-luminosity relation, and also compare magnitudes and colours of the    
GCs and UCDs with those obtained for a sample of Antlia dE,N nuclei and GCs.    
The discussion of the results and their set in the literature context  
is developed in Section\,4. Finally, a summary and the conclusions are provided in Section\,5.         
 
\section[]{Observations and reductions}    
 
\subsection{Photometric data and selection of point-sources}    
    
The photometric data set used in this paper consists of FORS1--VLT images 
from two fields in the $V$ and $I$ bands (programme 71.B-0122(A), PI B. Dirsch). 
One of them is centred on NGC\,3268, and the other is a comparison field, located 
$\approx 22'$ to the north-west direction. We also used wide--field ($36' \times 36'$) images 
that were taken with the MOSAIC camera mounted at the CTIO 4-m Blanco telescope 
during 4/5 April 2002. These images correspond to a single field from the central 
region of the Antllia cluster, including NGC\,3268, obtained in the Kron-Cousins 
$R$ and Washington $C$ filters.
For both data sets, the procedure applied was similar. First, the software SE\textsc{xtractor}  
\citep{ber96} was run on the images, once the extended galaxy light  
had been subtracted, to generate an initial catalogue of point-sources. Then  
the photometry was performed with \textsc{daophot} within \textsc{iraf} in the usual  
manner. The aperture  
photometry was carried out using the task PHOT. Afterwards, the definitive  
photometry was obtained with the task ALLSTAR, using a spatially variable  
point-spread function (PSF). The final point-source selection was based on  
the $\chi$ and sharpness parameters calculated by this task.

We estimated the photometry completeness for two FORS1--VLT fields, one of
them containing NGC\,3268, and the other one located to the north-west, at approximately
$15'$ from the gE galaxy. This latter field is going to be used as background region in
the rest of the paper. The procedure applied was the following. 
First, we added to each $V$ and $I$ image 1000 artificial stars with colours and 
magnitudes 
in the expected ranges for GCs. We repeated this process 10 times, in order to 
generate a sample of 10\,000 artificial stars, equally distributed over the entire
field. Then, we carried out the  photometry of these images in the same way
as the original ones. The completeness functions for the two fields are very similar. 
The $90\%$ completeness was achieved at $V\approx 24.5$, and the $60\%$ at 
$V\approx 25.75$.

In addition, two ACS fields (centred on each of the two Antlia dominant galaxies)  
observed with the $F814$\,filter were obtained with the  
HST (programme 9427; PI: W. E. Harris).  
Each image is the combination of four 570\,s exposures.  
Following a similar procedure to that applied in Paper\,I to the ACS image  
centred on NGC\,3258, we obtained the PSF of the other image centred on  
NGC\,3268, which was needed to measure $R_{\rm eff}$.  
With this purpose we applied SE\textsc{xtractor} on the galaxy-light  
subtracted image and obtained a point-source catalogue. Finally, the  
\textsc{iraf/daophot} tasks PHOT and PSF were applied to obtain the aperture  
photometry and PSF of the image, respectively. 
 
We refer to Paper\,I and references therein for details on the reduction and  
photometry of the three data sets (FORS1, MOSAIC, and ACS).  
 
From the objects that fulfill the point-source selection criteria  
applied to the FORS1 and MOSAIC data, we kept those with $-13.5 < M_V < -10.5$  
(see Paper\,I and references therein)  
and colours in the usual range for GCs \citep{dir03b,bas08}.  
Adopting for Antlia a distance modulus  
of $(m-M) = 32.73$ \citep{dir03b}, i.e. approximately 35\,Mpc, this magnitude  
range corresponds to $19.2 < V < 22.2$. In the following, we will refer to the  
sources in this preliminary selection as {\bf UCD candidates}.

\subsection{Spectroscopic data}    
 
We have also obtained GEMINI--GMOS multi-object spectra for objects located     
in  six Antlia fields (GS-2009A-Q-25, PI L. P. Bassino; GS-2010A-Q-21, PI L. P. 
Bassino; GS-2011A-Q-35, PI A. V. Smith Castelli; and GS-2013A-Q-37, PI J. P. Calder\'on). 
The masks designed for these programmes had slits not only on UCDs but also on other 
targets from GCs to galaxies. The grating B600\_G5303 blazed at $5000\,{\rm \mathrm{\AA}}$ 
was used, applying small shifts in the central wavelengths to fill the CCD gaps. The 
slit width was 1\,arcsec.   
The wavelength coverage for this configuration spans $3300 - 7200\,{\rm \mathrm{\AA}}$ 
(depending on the positions of the slits) and the resultant spectral resolution is 
$\sim 4.6\,{\rm \mathrm{\AA}}$. The total exposure times ranged between 2 and 3.5\,h. As part 
of the programme, we also obtained individual calibration flats and CuAr arc spectra 
for each exposure to correct for small variations that may be introduced by telescope 
flexion. Data reduction was performed using the GEMINI.GMOS package within \textsc{iraf} 
in the usual manner, following the same procedure explained in Paper\,I. For the faintest 
objects, it was not possible to trace  the individual exposures. In these cases,  
the exposures were combined to achieve a higher S/N. The trace of these objects were   
obtained from these combined images, and then used to extract the spectra from the  
individual exposures.     
 
We measured the heliocentric radial velocities for the UCD and GC candidates in the 
GMOS fields using the \textsc{iraf} task FXCOR within the NOAO.RV package. We used  
synthetic templates, selected from the single stellar population (SSP) model spectra 
at the MILES library (http://www.iac.es/proyecto/miles, \citealt{san06}). For this 
purpose, SSP models with metallicity [M/H]=\,-0.71, a unimodal initial mass function 
with slope $1.30$, and ages of 8 and 10\,Gyr were considered. The wavelength coverage 
of these templates is $4200 - 7300\,{\rm \mathrm{\AA}}$, and their spectral resolution is 
$2.3\,{\rm \mathrm{\AA}}$ FWHM. As in Paper\,I, the 10\,Gyr template provided slightly better 
correlations.   
 
\begin{table*}    
\begin{minipage}{110mm}    
\begin{center}    
\caption{Basic properties of the newly confirmed Antlia Compact Objects. They are  
labeled with the acronym `ACO' plus an order number, following the notation used in  
Paper\,I.}     
\label{table1}    
\begin{tabular}{@{}cc@{}c@{}c@{}c@{}cc@{}c@{}c@{}c@{}cccccr@{}c@{}l@{}}    
\hline    
\\    
\multicolumn{1}{c}{ID}&\multicolumn{5}{c}{RA(J2000)}&\multicolumn{5}{c}{DEC(J2000)}&\multicolumn{1}{c}{V$_0$}&   
\multicolumn{1}{c}{(V\,-\,I)$_0$}&\multicolumn{1}{c}{(T$_1$)$_0$} 
&\multicolumn{1}{c}{(C\,-\,T$_1$)$_0$}&\multicolumn{3}{c}{RV$_{\rm hel}$}\\    
\multicolumn{1}{c}{}&\multicolumn{5}{c}{hh mm ss}&\multicolumn{5}{c}{dd mm ss}&\multicolumn{1}{c}{mag}&   
\multicolumn{1}{c}{mag}&\multicolumn{1}{c}{mag}&\multicolumn{1}{c}{mag} 
&\multicolumn{3}{c}{km\,s$^{-1}$}\\    
    
\hline    
\\    
ACO\,12&$10$&$\,$&$29$&$\,$&$49.344$&$-35$&$\,$&$17$&$\,$&$40.128$&$21.19$&$1.03$&$20.68$&$1.79$&$2279$&$\pm$&$45$\\ 
ACO\,13&$10$&$\,$&$29$&$\,$&$52.584$&$-35$&$\,$&$19$&$\,$&$20.532$&$22.68$&$0.93$&$22.19$&$1.35$&$2668$&$\pm$&$44$\\ 
ACO\,14&$10$&$\,$&$29$&$\,$&$54.42$&$-35$&$\,$&$17$&$\,$&$47.472$&$22.48$&$1.13$&$21.91$&$1.90$&$2304$&$\pm$&$37$\\ 
ACO\,15&$10$&$\,$&$29$&$\,$&$54.888$&$-35$&$\,$&$18$&$\,$&$45.036$&$22.23$&$1.06$&$21.74$&$1.77$&$2470$&$\pm$&$24$\\ 
ACO\,16&$10$&$\,$&$29$&$\,$&$57.228$&$-35$&$\,$&$19$&$\,$&$28.416$&$21.33$&$1.07$&$20.81$&$1.60$&$2437$&$\pm$&$45$\\ 
ACO\,17&$10$&$\,$&$29$&$\,$&$58.704$&$-35$&$\,$&$21$&$\,$&$4.86$&$21.73$&$0.95$&$21.22$&$1.41$&$2860$&$\pm$&$18$\\ 
ACO\,18&$10$&$\,$&$29$&$\,$&$58.92$&$-35$&$\,$&$24$&$\,$&$8.028$&$-$&$-$&$20.67$&$1.76$&$2747$&$\pm$&$20$\\ 
ACO\,19&$10$&$\,$&$29$&$\,$&$59.316$&$-35$&$\,$&$18$&$\,$&$51.984$&$22.45$&$1.01$&$22.08$&$1.43$&$2865$&$\pm$&$32$\\ 
ACO\,20&$10$&$\,$&$30$&$\,$&$0.828$&$-35$&$\,$&$20$&$\,$&$20.076$&$21.87$&$0.99$&$21.41$&$1.53$&$3013$&$\pm$&$30$\\ 
ACO\,21&$10$&$\,$&$30$&$\,$&$1.764$&$-35$&$\,$&$20$&$\,$&$15.936$&$21.20$&$1.06$&$20.70$&$1.74$&$2628$&$\pm$&$16^1$\\ 
ACO\,22&$10$&$\,$&$30$&$\,$&$1.836$&$-35$&$\,$&$21$&$\,$&$21.6$&$22.58$&$1.10$&$22.02$&$1.85$&$2734$&$\pm$&$31$\\ 
ACO\,23&$10$&$\,$&$30$&$\,$&$1.908$&$-35$&$\,$&$20$&$\,$&$59.352$&$22.06$&$0.99$&$21.47$&$1.64$&$2277$&$\pm$&$36$\\ 
ACO\,24&$10$&$\,$&$30$&$\,$&$2.988$&$-35$&$\,$&$19$&$\,$&$10.02$&$21.76$&$1.04$&$21.28$&$1.64$&$2930$&$\pm$&$29$\\ 
ACO\,25&$10$&$\,$&$30$&$\,$&$3.132$&$-35$&$\,$&$20$&$\,$&$11.184$&$21.64$&$1.08$&$21.12$&$1.56$&$2796$&$\pm$&$32^1$\\ 
ACO\,26&$10$&$\,$&$30$&$\,$&$4.212$&$-35$&$\,$&$16$&$\,$&$18.048$&$-$&$-$&$21.55$&$1.93$&$2629$&$\pm$&$38$\\ 
ACO\,27&$10$&$\,$&$30$&$\,$&$4.356$&$-35$&$\,$&$20$&$\,$&$27.564$&$22.24$&$0.95$&$21.73$&$1.44$&$2712$&$\pm$&$19$\\ 
ACO\,28&$10$&$\,$&$30$&$\,$&$4.68$&$-35$&$\,$&$20$&$\,$&$2.868$&$22.72$&$1.09$&$22.10$&$1.64$&$2972$&$\pm$&$76$\\ 
ACO\,29&$10$&$\,$&$30$&$\,$&$5.436$&$-35$&$\,$&$21$&$\,$&$15.624$&$22.11$&$0.90$&$21.62$&$1.39$&$2611$&$\pm$&$40$\\ 
ACO\,30&$10$&$\,$&$30$&$\,$&$7.632$&$-35$&$\,$&$20$&$\,$&$51.432$&$21.68$&$1.03$&$21.17$&$1.62$&$2776$&$\pm$&$35$\\ 
ACO\,31&$10$&$\,$&$30$&$\,$&$7.884$&$-35$&$\,$&$16$&$\,$&$48.144$&$-$&$-$&$22.19$&$1.86$&$2890$&$\pm$&$34$\\ 
ACO\,32&$10$&$\,$&$30$&$\,$&$8.136$&$-35$&$\,$&$22$&$\,$&$40.512$&$21.58$&$0.92$&$21.11$&$1.85$&$2125$&$\pm$&$75$\\ 
ACO\,33&$10$&$\,$&$30$&$\,$&$9.828$&$-35$&$\,$&$17$&$\,$&$57.768$&$22.05$&$1.10$&$21.51$&$1.80$&$3200$&$\pm$&$22$\\ 
ACO\,34&$10$&$\,$&$30$&$\,$&$30.492$&$-35$&$\,$&$13$&$\,$&$2.604$&$-$&$-$&$20.51$&$1.76$&$2544$&$\pm$&$18$\\ 
ACO\,35&$10$&$\,$&$30$&$\,$&$30.511$&$-35$&$\,$&$36$&$\,$&$45.828$&$-$&$-$&$20.56$&$1.96$&$2660$&$\pm$&$43$\\ 
\hline    
\multicolumn{18}{l}{\small $^1$ ACOs were observed in two different programmes. The listed radial  
velocities are the weighted}\\ 
\multicolumn{18}{l}{ \small means of their individual measurements.}\\ 
\end{tabular}     
\end{center}     
\end{minipage}    
\end{table*}    
 
{\bf    
\subsection{Selected sample of GCs and UCDs}} 
    
According to our previous studies of the galaxy populations of the Antlia cluster,   
cluster members have radial velocities in the range $1200 - 4200$\,km\,s$^{-1}$  
\citep{smi08a,smi12}. Such a velocity range points to a complex structure, probably
due to a mixture of Hubble flow and internal cluster velocities.
Assuming the same criterion, we confirm as Antlia members 24 new objects
(23 of them in the vicinity of NGC\,3268, and the remaining one close to NGC\,3273),
listed in Table\,1. As the sample of newly confirmed objects include both, GCs and
UCDs (i.e., objects fainter and brighter than $M_V=-10.5$, respectively), we will
follow the notation adopted in Paper\,I, using the acronym `ACO' for  Antlia 
Compact Object.

Their J2000 coordinates, extinction corrected $V,I$ (when available) and Washington  
$C,T_1$ photometry, as well as heliocentric radial velocities are listed in this Table. 
Hereafter, extinction corrections are applied to magnitudes and colours. We refer to 
\citet{bas08} and \citet{dir03b} for explanations on how the extinction corrections 
applied to the $V,I$ and Washington $C,T_1$ data, respectively, have been calculated.

\section{Results}    
The mean radial velocity for our sample of 23 confirmed ACOs around NGC\,3268, is
$2720\pm56\,{\rm km\,s}^{-1}$. NED\footnote{This research has made use of the 
NASA/IPAC Extragalactic Database (NED) which is operated by the Jet Propulsion 
Laboratory, California Institute of Technology, under contract with the National 
Aeronautics and Space Administration.} lists for NGC\,3268 a radial velocity of 
$2800\pm21\,{\rm km\,s}^{-1}$. The agreement is acceptable, given the small object 
sample.

\begin{figure*}    
\includegraphics[width=85mm]{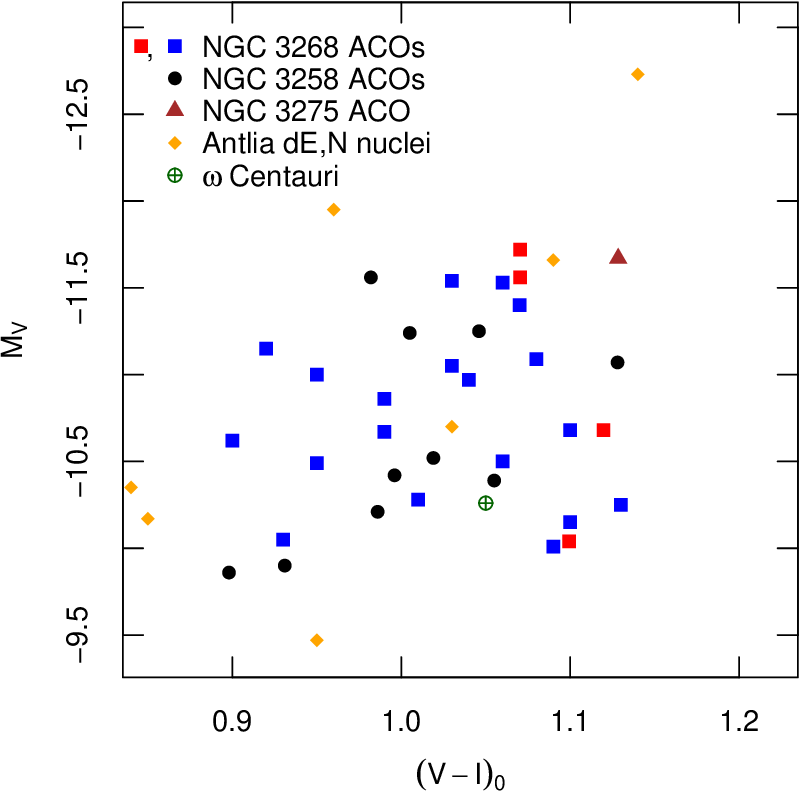}    
\includegraphics[width=84mm]{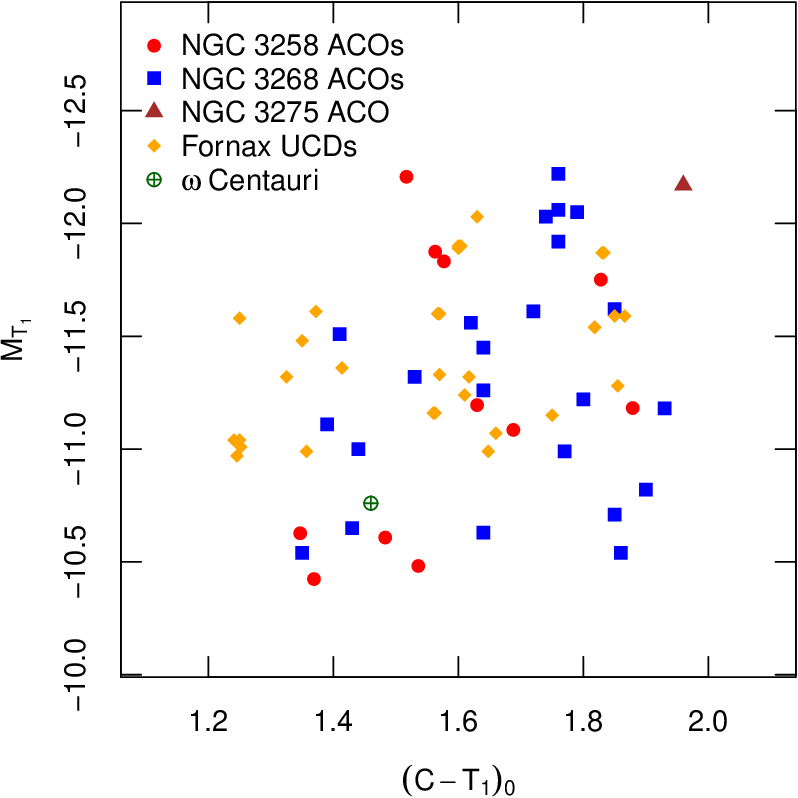}    
\caption{{\bf Left panel:} Colour-magnitude diagram of ACOs (confirmed objects) and $\omega$ Cen 
\citep[][2010 Edition]{har96} in the $(V,I)$ photometric system. The blue squares indicate 
the objects around NGC\,3268 with available $(V,I)$ photometry while for the red ones their 
colours and magnitudes were obtained from the Washington photometry (see text). Orange  
diamonds indicate the Antlia dE,N nuclei measured in Paper\,I. The brown triangle represents 
ACO\,35, the only object confirmed around NGC\,3273. Its colour and magnitude was also obtained 
from the Washington photometry. Black circles shows ACOs around NGC\,3258 (Paper\,I) with 
available $(V,I)$ photometry. 
{\bf Right panel:} Colour-magnitude diagram of ACOs in the Washington photometric system. The blue 
squares indicate the ACOs around NGC\,3268 with $(V,I)$ photometry available, while  the 
red circles represent the ACOs around NGC\,3258 (Paper\,I). Confirmed Fornax UCDs from
\citet{mie04} with available Washington photometry \citep{dir03a,bas06a}  and similar 
luminosities are shown with orange diamonds. $\omega$ Cent \citep[][2010 Edition]{har96} is 
also shown.}
\label{cmd}    
\end{figure*}    

{\bf
\subsection{Colour-magnitude diagram of ACOs and dwarf elliptical nuclei}}
\label{sec_cmd}    
Only those ACOs located in the VLT fields have $(V,I)$ photometry available. For the rest 
of them, we attempted to transform Washington photometry to the $V$ Johnson and $I$ Cousins 
system. To this aim, we calculated $V-R$ for the GC-like sources in the vicinity of NGC\,3268 
that have photometry in both photometric systems, i.e. in common between MOSAIC and VLT 
samples. Removing outliers iteratively, we obtained a mean and dispersion of 
$(V-T_1)_0 = 0.51\pm0.01$ and $\sigma_{(V-R)_0} = 0.09$. This value is in agreement with 
$V-R= 0.5$ used by \citet{mie04} in Fornax \citep[if we assume $R-T_1=0.02$, from][]{dir03b}. 
We also search for a possible correlation between $(V-I)_0$ and $(C-T1)_0$ colours. We 
found that a straight line gives an acceptable fit, resulting
$(V-I)_0 = 0.29\pm0.01 \times (C-T_1)_0 + 0.56\pm0.02$. This relation, together with 
$(V-R)_0 = 0.51$, will be used hereafter to transform the Washington photometry of those 
objects lacking $(V,I)$ data.
However, there is a large scatter, particularly for the colours in the usual range of 
{\it bona fide} metal-poor GCs. Objects with $0.8<(V-I)_0<0.9$ span $1.1<(C-T_1)_0<1.8$. 
For this reason, the transformed $V,I$ colours should be considered with caution. We use 
them only for comparison with the literature.

The $(V,I)$ CMD of the ACOs is shown in the left panel of Fig.\,\ref{cmd}.  
The blue squares indicate the ACOs around NGC\,3268 with available $(V,I)$ 
photometry while the red squares identify those ACOs whose colours and magnitudes 
were obtained from the Washington photometry, as described above. ACO\,35, the 
only object measured around NGC\,3273, is shown with a brown triangle, and the 
ACOs associated with NGC\,3258 listed in Paper\,I with black circles. Orange 
diamonds represent the nuclei of dE,N galaxies that are members of the Antlia cluster  
and were studied in Paper\,I.
The right panel of Fig.\,\ref{cmd} shows the Washington CMD for the ACOs around
NGC\,3258 (red circles, Paper I), NGC\,3268 (blue squares) and NGC\,3273 (brown triangle).
A sample of confirmed Fornax UCDs from \citet{mie04} with available Washington photometry 
\citep{dir03a,bas06a} and similar luminosities is also indicated (orange diamonds).
For comparison, the position of $\omega$ Cent \citep[][2010 Edition]{har96} is 
indicated in both CMDs.
  
None of the ACOs confirmed in this paper nor in the previous one present a luminosity      
as high as those of the brightest Virgo or Fornax UCDs \citep[e.g.,][]{gre09,bro11}.  
Blue ACOs around NGC\,3258 and NGC\,3268 seem to occupy the same locus in the 
CMD as the nuclei of Antlia dE,N, that follow a colour-luminosity correlation, getting 
redder when their luminosities increases (Paper\,I and references therein). Considering that our ACOs are
mainly fainter that $M_V=-11.5$, blue ones are more likely to have a common origin
with dE,N nuclei than red ones.

Out of the sixteen ACOs in the neighbourhood of NGC\,3268 brighter than $M_V = -10.5$, 
seven are redder than $V-I = 1.05$, a usual limit between metal-poor (`blue') and  
metal-rich (`red') GCs \citep[e.g.,][]{bas08}. In comparison with NGC\,3258 ACOs, this  
represents a higher fraction of red ACOs.

The picture is quite different if we analyse the Washington colours. Just four ACOs
brighter than $M_{T1} = -11$ are bluer than $C-T_1 = 1.55$ a value commonly used as
the limit between metal-poor and metal-rich GCs in this photometric system 
\citep[e.g.,][]{dir03a,dir03b,dir05,bas06a,bas06b}, and all of them are redder than 
$C-T_1 = 1.4$. Fornax UCDs also present a large fraction of UCDs with red
colours, but the sample spreads over a larger colour range, with some bright objects 
presenting blue colours.

\subsection{Effective radii of ACOs} 
The high resolution of the HST data allows us to measure effective radii ($R_{\rm eff}$)  
of GCs or UCDs \citep[e.g.][]{mie07,mie08,evs08,mad10,cas13a} at distances as large as 
that of the Antlia cluster. At the adopted Antlia distance, the ACS pixel size of 
0.055\,arcsec corresponds to $\sim\,9.3$\,pc. Sizes of a similar order have been obtained 
for a large fraction of UCDs \citep{mie08,bro11,mis11}.   
 
\textsc{ishape} \citep{lar99} was used to fit the light profiles of ACOs located  
in the NGC\,3268 ACS field and obtain their $R_{\rm eff}$. We chose a King profile  
\citep{kin62,kin66} with concentration parameter $c=30$ (defining $c$ as the ratio 
of the tidal over the core radius). King30-profiles have been  used in Paper\,I 
as well as in  previous work \citep[e.g.][]{har09,mad10,bro11}.
 We refer to Paper\,I for more details on how \textsc{ishape} was run. 
    
Twelve ACOs spectroscopically  
confirmed in this paper are located within the NGC\,3268 ACS field. In  
all the cases, we measured $2.8 < R_{\rm eff} [{\rm pc}] < 8.5$. 
For testing the reliability of the fit, we applied \textsc{ishape} to seven   
foreground stars confirmed with radial velocities, and obtained typically
FWHM-values of $\sim 0.01-0.03$\,pixels, that is approximately one tenth of 
the smallest FWHM obtained for the confirmed ACOs.  
We also measured $R_{\rm eff}$ for objects classified as point-like sources 
from the MOSAIC and FORS1 photometry, that were located in the two ACS fields, 
and present colours in the same range as GCs and $M_V < -9$, i.e. we added fainter 
GC candidates to our sample.
This magnitude limit was adopted so as to reach  $S/N >70$. \citet{lar99} recommends 
an $S/N >50$ for obtaining accurate shape parameters with \textsc{ishape}.
 
\begin{figure}    
\includegraphics[width=84mm]{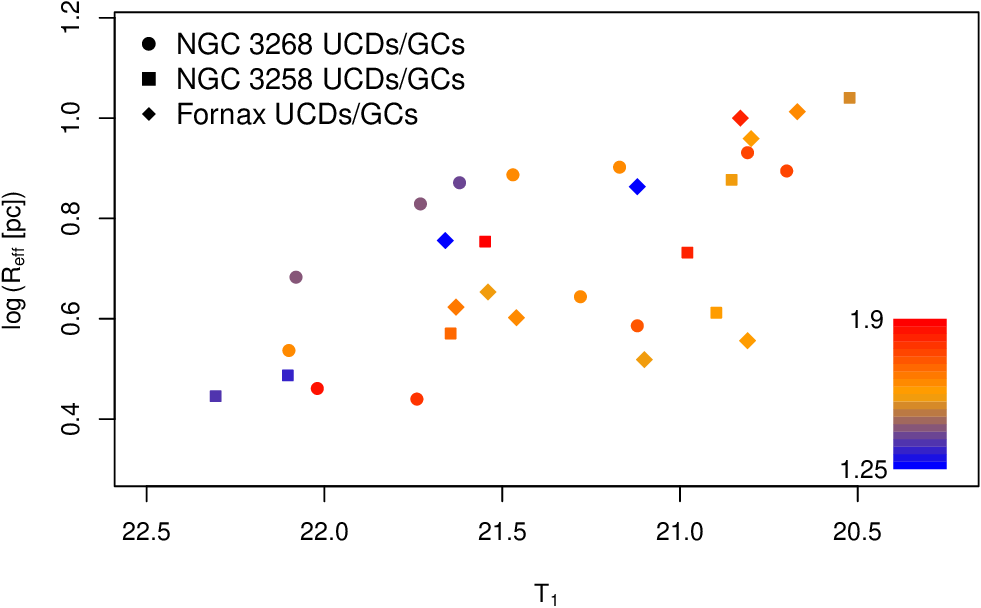}   
\caption{${\rm log}(R_{\rm eff})$ vs. {\bf $T_1$} for Antlia GCs ($T_1 > 21.6$) and 
UCDs ($T_1 < 21.6$) located close to NGC\,3268 in the ACS field (circles), and the 
ACOs from Paper\,I whose $R_{\rm eff}$ was measured (squares), plus Fornax UCDs from 
\citet{mie08}
with available Washington photometry \citep{dir03a,bas06a} and similar
luminosity (diamonds), converted to the Antlia distance. The color palette 
represents the Wasington colour for the objects, spanning $1.25 < (C-T_1)_0 < 1.9$.}    
\label{reff1}    
\end{figure}    
   
As explained in Paper\,I, the FWHM of the ACOs analyzed with \textsc{ishape} is 
less than one pixel, so possible  eccentricities will not be considered, but the 
ellipticities of UCDs or GCs are anyway not large (e.g. \citealt{har09} and 
references therein; \citealt{chib11}). 
   
Fig.\,\ref{reff1} shows ${\rm log}(R_{\rm eff})$ vs. {\bf $T_1$}, for all confirmed ACOs 
in the NGC\,3258 and NGC\,3268 ACS fields. ACOs identified in Paper\,I are  
indicated with squares, while  circles are used for the new ACOs presented in this  
paper. The $R_{\rm eff}$ for the ACOs around both Antlia gEs are similar,
 presenting an upper limit of approximately 10\,pc. Diamonds represent
Fornax UCDs from \citet{mie08} with available Washington photometry 
\citep{dir03a,bas06a} and similar luminosity, converted to Antlia distance. 
Considering the Antlia samples together, a $R_{\rm eff}-M_{T1}$ trend could exist.

\begin{figure}    
\includegraphics[width=84mm]{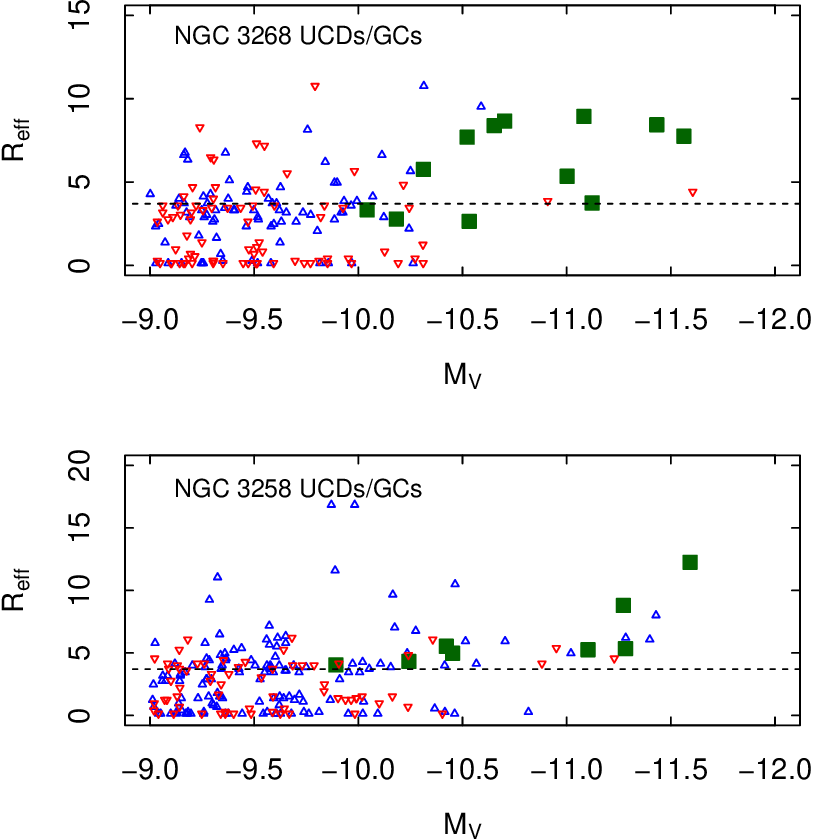}    
\caption{{\bf Top panel:} $R_{\rm eff}$ vs. $M_V$ for the GC and UCD candidates around  
NGC\,3268, discriminated between those objects bluer and redder than $(V-I)_0=1.05$ (blue  
triangles and red inverted triangles, respectively). Confirmed ACOs in the vicinity of this  
galaxy are indicated as green filled symbols.}.  
{\bf Bottom panel:}
 Analogue for the candidates  
around NGC\,3258. The dashed line in both panels is the mean $R_{\rm eff}$ for a sample of  
Galactic GCs \citep[][2010 Edition]{har96}.    
\label{reff2}    
\end{figure}

The top panel of Fig.\,\ref{reff2} shows the $R_{\rm eff}$ vs. $M_V$ for UCDs and GCs candidates  
brighter than $M_V = -9$ around NGC\,3268, with colours in the same range as  
GCs. As explained above, the value  
$(V-I)_0=1.05$ is adopted to separate between metal-poor and metal-rich GC 
candidates, which are indicated with blue triangles and red inverted triangles, respectively. 
Spectroscopically confirmed ACOs are shown with green filled squares. 

The dashed line represents the mean $R_{\rm eff}$ for a sample of 84 Galactic GCs with $-10 < M_V < -7$  
\citep[][2010 Edition]{har96}, which results in $3.7\pm0.3$\,pc. For our sample, we calculate  
the mean $R_{\rm eff}$ for the blue and red candidates fainter than $M_V\approx-10.5$. 
Both subsamples are restricted to objects with $1 < R_{\rm eff}\,{\rm [pc]} < 10$. This lower limit  
was chosen because at the Antlia distance, smaller $R_{\rm eff}$ are very uncertain due to the  
\textsc{ishape} fitting limitations \citep{har09}.  
The upper limit leaves the ECs out of the sample, as these objects present masses and luminosities 
comparable with GCs, but effective radii larger than 10\,pc \citep{bru12}.

The resulting $R_{\rm eff}$ averages for blue and red GC candidates are $3.4\pm0.15$ 
and $3.3\pm0.1$\,pc, respectively. The bottom panel is analogous, but for the candidates around 
NGC\,3258, and their mean $R_{\rm eff}$-values are $3.4\pm0.15$ and $3.1\pm0.2$\,pc,  
respectively. In both cases, the differences in the mean $R_{\rm eff}$ for blue and red GCs   
are within the uncertainties. This results is in agreement with \citet{nan11} who found 
that M\,81 blue and red GC candidates present nearly identical half-light radius, despite that
a difference between blue and red GCs has benn found in other systems \citep[e.g.][]{lar01b}.
They mean $R_{\rm eff}$ for blue and red GCs in our samples are in agreement, within the errors, 
with the mean value obtained for the Galactic GCs. 
 
We adopted the same limit as the one proposed in Paper\,I, i.e. $M_V = -10.5$, to 
discriminate between `regular' GCs and brighter objects. As a consequence, the 
small open triangles with $M_V < -10.5$ in that figure will be considered as UCD 
candidates.  
 
Still in Fig.\,\ref{reff2}, we also find GC candidates fainter than $M_V = -10.5$  
around both gEs with $R_{\rm eff} \gtrapprox 10$\,pc, that appear separated    
from the bulk of GC candidates. Thus, following the literature definition 
\citep[][and references therein]{mad11,bru12}, we classified them as EC 
candidates. There are in total nine EC candidates, two of the around NGC\,3268 
and seven near NGC\,3258. $(V,I)$ is available for all the ECs, but Washington
one is available for just eight of them. Six have colours resembling blue GCs 
in both photometric systems, and only one is a red GC considering $(V,I)$ and
Washington. The coordinates, magnitudes and colours, and $R_{\rm eff}$ for
the EC candidates are indicated in Table\,\ref{table2}.
The two largest EC candidates are located around NGC\,3258, with 
$R_{\rm eff}\approx 17$\,pc. Both objects have $(V-I)_0$ colours in the range  
of blue GCs, $M_V \approx -10$ and are located at $\sim 30\arcsec$ from NGC\,3258 
centre.

\begin{table*}    
\begin{minipage}{110mm}    
\begin{center}    
\caption{Basic properties of the extended cluster candidates (defined as
objects with $M_V \gtrapprox -10.5$ and $R_{\rm eff} \gtrapprox 10$\,pc)
around both giant ellipticals.}     
\label{table2}    
\begin{tabular}{@{}c@{}c@{}c@{}c@{}cc@{}c@{}c@{}c@{}cccccr@{}c@{}l@{}}    
\hline    
\\    
\multicolumn{5}{c}{RA(J2000)}&\multicolumn{5}{c}{DEC(J2000)}&\multicolumn{1}{c}{V$_0$}&   
\multicolumn{1}{c}{(V\,-\,I)$_0$}&\multicolumn{1}{c}{(T$_1$)$_0$} 
&\multicolumn{1}{c}{(C\,-\,T$_1$)$_0$}&\multicolumn{3}{c}{$R_{\rm eff}$}\\    
\multicolumn{5}{c}{hh mm ss}&\multicolumn{5}{c}{dd mm ss}&\multicolumn{1}{c}{mag}&   
\multicolumn{1}{c}{mag}&\multicolumn{1}{c}{mag}&\multicolumn{1}{c}{mag} 
&\multicolumn{3}{c}{pc}\\    
\hline    
\\
$10$&$\,$&$30$&$\,$&$10.22$&$-35$&$\,$&$19$&$\,$&$42.66$&$22.42$&$0.93$&$21.94$&$1.33$&$10.74$\\
$10$&$\,$&$29$&$\,$&$58.87$&$-35$&$\,$&$19$&$\,$&$02.54$&$22.94$&$1.11$&$22.26$&$1.88$&$10.60$\\
$10$&$\,$&$28$&$\,$&$55.40$&$-35$&$\,$&$36$&$\,$&$23.43$&$22.86$&$0.93$&$-$&$-$&$16.80$\\
$10$&$\,$&$28$&$\,$&$55.94$&$-35$&$\,$&$36$&$\,$&$31.71$&$22.84$&$0.89$&$22.10$&$1.47$&$11.56$\\
$10$&$\,$&$28$&$\,$&$52.49$&$-35$&$\,$&$35$&$\,$&$56.54$&$23.4$&$0.92$&$23.13$&$1.80$&$9.22$\\
$10$&$\,$&$28$&$\,$&$47.24$&$-35$&$\,$&$35$&$\,$&$37.27$&$23.4$&$0.91$&$22.88$&$1.49$&$11.01$\\
$10$&$\,$&$28$&$\,$&$49.06$&$-35$&$\,$&$35$&$\,$&$26.44$&$22.56$&$0.97$&$21.94$&$1.42$&$9.63$\\
$10$&$\,$&$28$&$\,$&$51.94$&$-35$&$\,$&$34$&$\,$&$58.02$&$22.27$&$0.98$&$21.62$&$1.50$&$10.46$\\
$10$&$\,$&$28$&$\,$&$56.64$&$-35$&$\,$&$36$&$\,$&$08.81$&$22.75$&$0.97$&$22.10$&$1.52$&$16.79$\\
\hline    
\\
\end{tabular}     
\end{center}     
\end{minipage}    
\end{table*}

\subsection{Comparison with the blue tilt extrapolation}  
 
\begin{figure}    
\includegraphics[width=84mm]{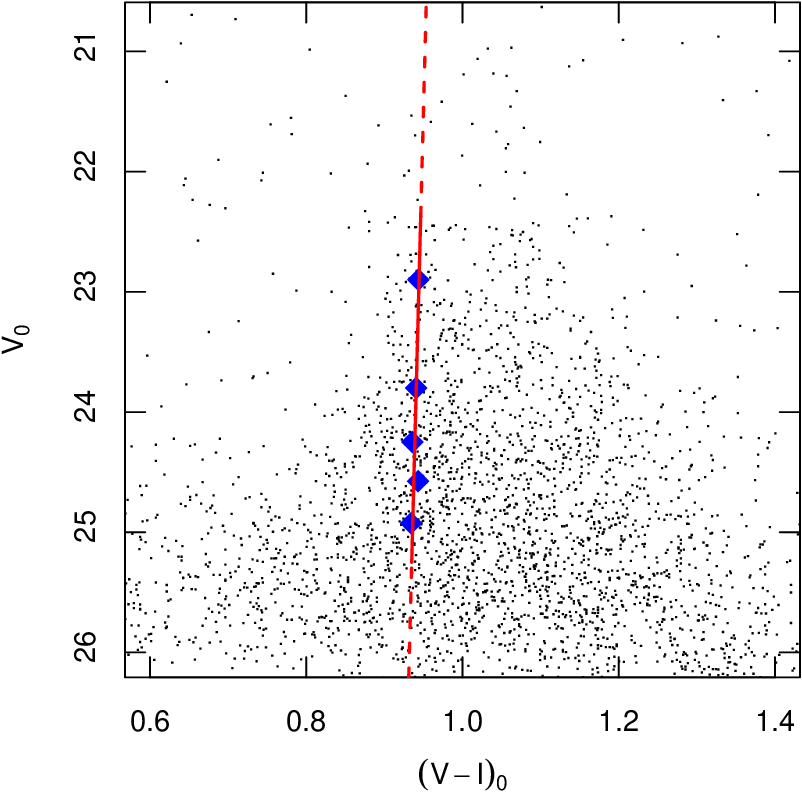}   
\caption{$(V,I)$ CMD for the NGC\,3268 GCS. Filled blue diamonds are the mean colours of blue GCs in  
five equally populated magnitude bins. The red solid line represents the fit to these points, while  
the dashed lines indicate its extrapolation.}    
\label{cmd_bt}    
\end{figure}

In this section we first investigate  whether the so-called `blue tilt' is present in the GCS of NGC\,3268,  
as it is in NGC\,3258, and then compare its extrapolation with the colours of ACOs and UCDs candidates.  
In some galaxies, a correlation that has been named `blue tilt' is followed by the blue GCs in  
the CMD, in the sense that bright blue GCs get redder when we move towards brighter luminosities.  
It is understood as a mass-metallicity relation for this metal-poor subpopulation of old GCs.  
This effect had been noticed in the Antlia gEs \citep{har06}. In Paper\,I we calculated a slope
of $d(V-I)_0/dV_0 \approx -0.03$ for NGC\,3258 GCS. Now, we take into account the blue GC candidates 
in the FORS1 field that contains NGC\,3268 with $22.3<V_0<25.75$ and $0.8(V-I)_0<1.05$. 
The background sample contains the point-sources that fulfill the same criteria than the blue GC 
candidates in a FORS1 field located towards the north-east from NGC\,3268 \citep[see Fig.\,1 in][]{bas08}. 
In order to detect if a blue tilt is also present in the NGC\,3268 GCS, we statistically subtracted the 
background contribution.  
For this purpose, we separated into cells the CMD of the GC candidates and the background samples. 
Then, we counted the number of objects found in each cell of the background sample, and subtracted  
randomly the same number of GC candidates in the corresponding cell. Afterwards, the GC  
candidates from the clean sample were split into five equally populated magnitude bins,  
spanning in the range $22.3<V_0<25$. For each bin we calculated the mean, that are represented  
by filled diamonds in the CMD of NGC\,3268 GCS (Fig.\,\ref{cmd_bt}). The solid line was obtained  
by fitting the mean values, while the dashed lines indicate its extrapolation towards brighter  
and fainter magnitudes. We obtained a slope $d(V-I)_0/dV_0 \approx -0.004$, which at most indicates  
just a marginal effect. Hence, a blue tilt in the $V$ vs. $(V-I)$ CMD is detected for the blue GCs  
around NGC\,3258 but not for NGC\,3268. On the other hand, its presence can be seen for both galaxies  
in the respective $I$ vs. $(B-I)$ CMDs performed by \citet{har06} with ACS data. 

\begin{figure}    
\includegraphics[width=84mm]{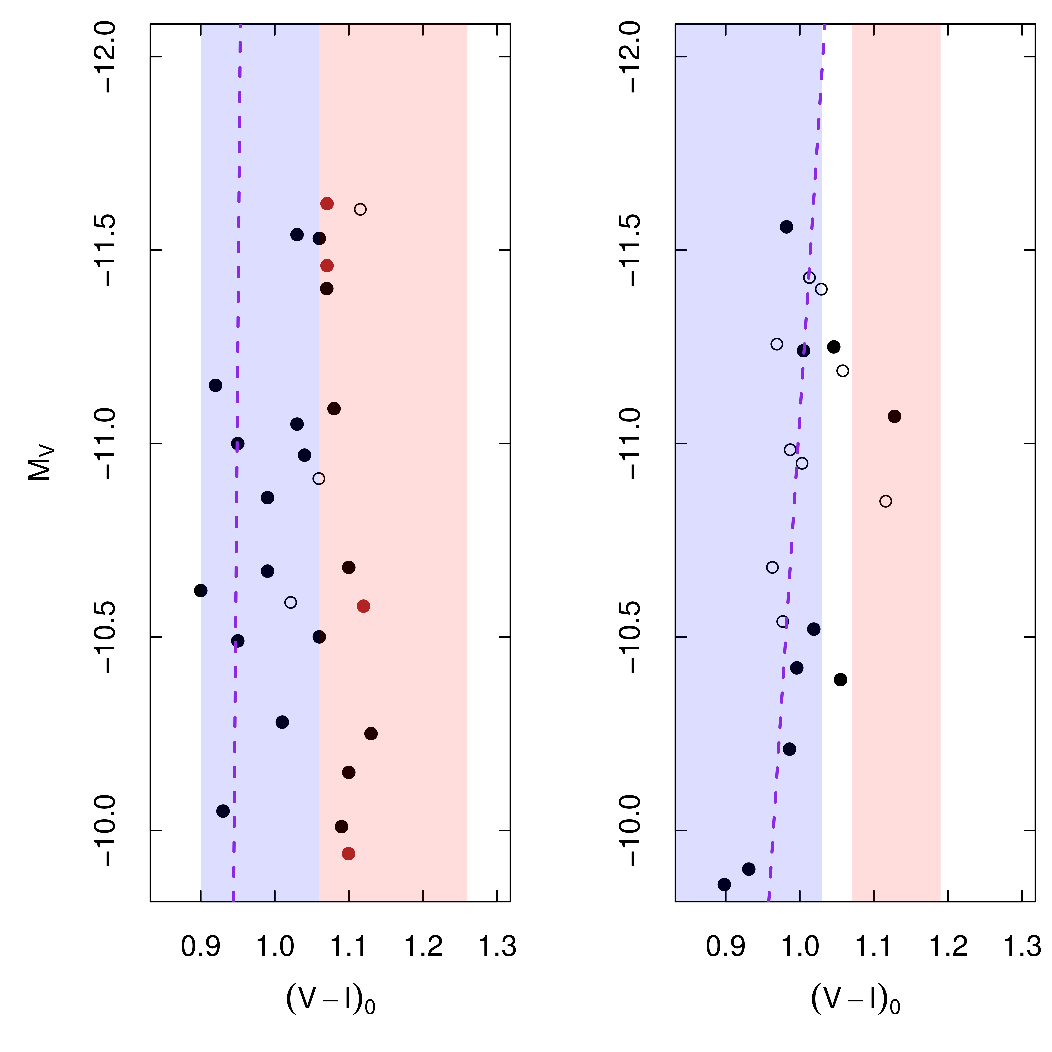}   
\caption{{\bf Left panel:} $(V,I)$ CMD for the NGC\,3268 ACOs (filled circles)  
and UCD candidates (open circles). The black circles indicate ACOs with available  
$(V,I)$ photometry while the brown ones identify ACOs with $(V,I)$ photometry  
derived from the Washington photometry (see Sect.\,\ref{sec_cmd}). The dashed  
line indicates the extrapolation of the `blue tilt'. The blue and red regions  
are centred on the peaks of the Gaussians fitted to the blue and red GCs colour  
distributions, respectively, while their widths are twice the respective dispersions   
\citep[values from ][]{bas08}. {\bf Right panel:} As in the left panel but for the NGC\,3258  
ACOs and UCD candidates from Paper\,I.}    
\label{cmd_sep}    
\end{figure}

The left panel of Fig.\,\ref{cmd_sep} shows the $(V,I)$ CMD for the NGC\,3268 ACOs  
 (filled circles) and UCD candidates (open circles).  
The black circles indicate the ACOs with available $(V,I)$ photometry and the brown circles  
the ones located outside the FORS1 fields, for which such photometry was derived from the  
$(C,T_1)$ Washington photometry (see Sect\,\ref{sec_cmd}). The dashed  
line indicates the extrapolation of the `blue tilt'. The blue and red regions in the CMD  
are centred on the $(V-I)$ peaks of the Gaussians fitted to the blue and red GC colour  
distributions, respectively, and their widths are twice the corresponding dispersions  
\citep[values from ][]{bas08}. The right panel is like the left panel but for the NGC\,3258  
ACOs and UCD candidates presented in Paper\,I. 
 
Fig.\,\ref{cmd_sep} shows that, as already mentioned in Paper\,I, ACOs and UCD candidates around  
NGC\,3258 (right panel) seem to follow the extrapolation of the blue tilt towards brighter  
luminosities. However, this behaviour is not similar in NGC\,3268 (left panel). In this latter  
galaxy, the brightest ACOs (with $M_V \approx -11.5$) have colours close to the limit between blue  
and red GCs, i.e. $(V-I)_0 \sim 1.05$. At the same time, the subsample corresponding to the rest  
of compact objects with fainter luminosities occupy a wider colour range, including the whole range  
of blue GCs (two thirds of the subsample) and the bluest side of the range of red GCs (one third  
of the subsample). It is also clearly seen that, as mentioned in Sect\,\ref{sec_cmd},  
the fraction of ACOs with colours like red GCs is higher in NGC\,3268 than in NGC\,3258, and the  
same applies to the UCD candidates.

\subsection{Expected number of bright GCs around NGC\,3268}  

In the following we will compare the number of bright GCs, estimated from the luminosity  
function of the GC candidates (GCLFs), with the actual number of ACOs and UCD candidates we 
have identified in the ACS field. 

We obtained the GCLF around NGC\,3268 from our FORS1 photometry,
taking into account GC candidates with $0.7 < (V-I)_0 < 1.3$ in the radial regime 
$0.33\arcmin < R_{proj} < 2\arcmin$. The LF was corrected for incompleteness and background 
subtracted. 
The GCLF is usually represented by a Gaussian profile \citep[e.g.][]{bas08} or a t5 function 
\citep[e.g.,][]{har01,ric03}. We fitted both profiles to the GC candidates brighter than 
$V=25.75$ (which corresponds to a completeness of $\approx 60\%$). For the Gaussian profile, 
the fitted scalling factor, maximum magnitude and dispersion were $547\pm15$\,GCs, $25.2\pm0.1$\,mag and 
$1.16\pm0.08$\,mag, respectively. In the case of the t5 function, the 
scalling factor and maximum magnitude result $566\pm15$\,GCs and $25.1\pm0.04$\,mag.

The numerical integration for GCs brighter than $M_V = -10.5$, within $2\arcmin$ from 
the centre of NGC\,3268 (this area is similar to the ACS field of view), gives $8\pm3$ 
or  $25\pm1$ bright GCs, using Gaussian and t5 profiles, respectively. Our 
previous analysis of the ACS field centred on NGC\,3268, showed that it contains 9 ACOs 
plus 3 UCD candidates, all objects brighter than $M_V = -10.5$. This falls within the 
range of the number of bright GCs estimated from both profiles.
We can extrapolate the previous results out to $10\arcmin$ from the centre of NGC\,3268,  
obtaining $22\pm8$ and $70\pm3$ GCs brighter than $M_V = -10.5$, using the Gaussian and  
the t5 profiles, respectively. 

We could try a different approach. Assuming for the GCS of NGC\,3268 a population of 4750 members  
\citep{bas08}, we generate 100 samples of the same size with a Monte-Carlo code. We use as   
distribution function the Gaussian profile fitted previously. For different Monte-Carlo runs,  
we perform slight variations in the distribution parameters, asuming a normal distribution for
their errors. Then, we find that the mean magnitude of the brightest object  
is $M_V\approx-11.7$, with a high scatter. In less than $10\%$ of the cases the brightest 
object reaches $M_V\approx-13$. Afterwards, we look for the
magnitudes of the five brightest objects in each run. The mean magnitude of the fifth brightest  
artificial GCs is $M_V\approx-11.1$. These results suggest  
that the magnitudes of ACOs around NGC\,3268 are within the expected values for the bright end of the 
GCLF. This was also found when we generated the samples from the t5 distribution.
 
\subsection{Mass estimate for NGC\,3268}
\label{sect.mass}
The sample of radial velocities is much too small to perform a radially 
dependent analysis of the mass profile, but some mass estimators have been 
proposed whose strengths it is to make best use of such small samples.

We employ the "tracer mass estimator" of \citet{eva03} which estimates the 
mass enclosed by the outermost GC by

\begin{equation}
 M =    \frac{C}{G N}  \sum_i V^2_{LOS,i} R_i 
\end{equation}

\noindent where $R_i$ and $V_{LOS,i}$ are the projected distances from NGC 3268 
and velocities relative to the mean velocity of the sample, respectively. G is the constant of 
gravitation, N the number of tracers and the constant C is calculated in the case
of isotropy through:

\begin{equation}
C = \frac{ 4\,(\alpha + \gamma)}{\pi} \frac{4 - \alpha - \gamma}{3- \gamma}
\frac{1-(r_{in}/rÌ£_{out})^{3-\gamma}}{1-(r_{in}/rÌ£_{out})^{4-\alpha-\gamma}}
\end{equation}

One further assumes that the three-dimensional tracer population profile obeys 
a power law between an inner radius $r_{in}$ and and outer radius $r_{out}$, as 
well as does the gravity. The corresponding exponents are $\gamma$ and $\alpha$, 
respectively. We adopt $\gamma=2.4$ for NGC 3268 from \citet{dir03b} and assume 
a constant circular velocity, which means $\alpha = 0$.
We assume that $r_{in}$ and $r_{out}$ are the projected distance to NGC\,3268 of
the innermost and outermost data point, respectively.

Then we get for the constant C=6.54. We exclude from the analysis the outermost 
ACO, because its projected distance to NGC\,3268 is more than twice the projected 
distance for the rest of them. Then, the mass within 47.2 kpc is 
$2.7\times 10^{12} {\rm M_\odot}$.

\section{Discussion} 

\citet{mie12} address two possible formation channels for the UCDs: that they are tidally disrupted 
dwarf galaxies or just the brightest members of the GC population.   
They perform a statistical analysis based on the sizes of the UCD samples in different environments  
and conclude that the latter option, i.e. UCDs being massive GCs, can fully explain the number of  
discovered UCDs. Our results support this conclusion, despite the differences in the estimated number
of GCs for gaussian and $\rm{t_5}$ functions.

The galaxy `threshing' scenario, i.e., the tidal disruption of a dE,N galaxy where only remains its 
nucleus, has been proposed by several authors \citep{bas94,bek01,bek03a,goe08}. Moreover, it has  
been suggested as the possible origin for $\omega$\,Cent \citep{lee99,bek03b,wyl10}, and other  
remarkable extragalactic compact objects \citep[e.g.][]{str13}. 
The simulations made by \citet{pfe13} show that, for a galaxy with a nucleus of $M_V = -10$, 
depending on the dwarf galaxy structure and its orbital parameters, an object as faint as
 $M_V= -9 - -10.2$ and a few parsecs of half-light radius could be generated. This implies that 
the remnant of disrupted dE,N galaxies could present similar photometric properties to many of the 
UCDs and bright GCs in our sample.
The UCDs in our sample ocupy a similar region in the CMD than Antlia dE,N nuclei, particularly for the
bluer objects. 

The mass estimator applied to the heliocentric radial velocities of the ACOs presented in this work 
indicates that NGC\,3268 is a massive elliptical galaxy. Its mass up to $\approx 47\,{\rm kpc}$ is
similar than that of many giant ellipticals in dense environments \citep[e.g.][]{hum06,sch10,sch12}.

Although new XMM-Newton observations exist, constraints of the dynamical mass still come from ASCA 
data \citep{nak00}. They describe the X-ray emission around NGC 3268 as isothermal with $kT_e=2\,{\rm keV}$. 
From their beta model of the X-ray surface brightness, 

\begin{equation}
n(r) = n_0 (1+(r/r_c)^2)^{-3 \beta/2}
\end{equation}

\noindent we adopt $\beta=0.38$ and $r_c=5'$.
Then, under the asumption of spherical symmetry, we can apply the formula \citep{gre01} 

\begin{equation}
M(r) =  \frac{3 \beta k T_e}{G \mu m_p} \frac{r^3}{r_c^2+r^2}
\end{equation}

where G is the constant of gravitation, $\mu$ the molecular weight, and $m_p$ the proton mass. We 
adopt $\mu=0.6$ and get for the mass within the volume defined by the globulars ($47.2\,{\rm kpc}$) 
$1.8\times 10^{12}$ which is in reasonable agreement with the mass derived from GCs.

Considering the luminosity profile and the mean colour $V-R \approx 0.7$ derived for NGC\,3268 by
\citet{dir03b}, its integrated magnitude up to $\approx 47\,{\rm kpc}$ is $V \approx 10.65$, which 
implies (assuming $m-M = 32.73$) a total mass-to-light ratio of $\Upsilon_{\odot} \approx 46.5$. This
value would indicate the presence of a massive group-scale halo.

\section{Summary and conclusions}    

Using CTIO (MOSAIC\,II), VLT (FORS1), and archival HST (ACS) imaging data  
as well as GEMINI (GMOS) spectroscopic data we have studied a sample
of objects in the Antlia cluster, that could be identified as UCDs and
bright GCs. We discuss the entire sample of UCDs identified until now 
in the Antlia cluster, including those around NGC\,3258 (Paper\,I).
We summarize our results and conclusions as follows. 
 
\begin{itemize} 
 
\item We present 24 objects in the Antlia cluster that, according to their  
luminosities, are classified either as bright GCs or as UCDs. From the 
total, 23 are located around one of the two central Antlia galaxies, 
NGC\,3268. The whole sample of "Antlia Compact Objects" (ACO) now consists 
of 35 spectroscopically confirmed objects. 
 
\item For a subsample of ACOs, UCD candidates and GC candidates located 
within the ACS fields, $R_{\rm eff}$'s have been  measured with the  \textsc{ishape} 
software. The GC candidates in the magnitude range $-9 < M_V< -10.5$ seem to 
have almost equal effective radii, whose mean value is in agreement with the 
mean value of  Galactic GCs. For objects brighter than $M_V = -10.5$,
classified as UCD candidates, a size-luminosity relation could exist, but 
there is no compelling evidence. The UCDs in this paper do not present
$R_{\rm eff}$ as large as Virgo ones.
 
\item In a $M_V$ vs. $(V-I)$ CMD, blue ACOs cover the same colour and magnitude 
ranges as the nuclei of dE,N galaxies studied in Paper\,I. Some of the ACOs may 
thus be remnants of nucleated galaxies. In comparison with Fornax, we 
did not find any UCD as bright as $M_V\approx-13$. The ACOs in our sample
present similar brightnesses, colours and radii as faint UCDs in Fornax. 
A comparison with Virgo UCDs could be biased due to the selection limit in
$R_{\rm eff}$ applied by the authors in that case.

\item We have discovered 10 objects around both central galaxies, with magnitudes  
similar to GCs, but having larger radii ($10 < R_{\rm eff} < 17$\,pc). They appear 
to be similar to the `extended clusters' found in the Milky Way and other 
galaxies \citep[][and references therein]{mad11,bru12} but that still have to be 
confirmed as members. Most of them have colours similar to blue GCs.
 
\item From the GCLF of NGC\,3258 and NGC\,3268 we estimated the expected number 
of GCs brighter than $M_V=-10.5$. These values, once corrected by the areal coverage 
of our UCDs search, are of the same order as our UCD samples. In this case, it would 
be not necessary to invoke a different origin to explain the number of UCDs.
Finally, in the scheme proposed by \citet{nor11}, our results support the idea that 
UCDs could be both, previous galaxy nuclei or normal, but bright GCs.

\item We used the sample of spectroscopically confirmed ACOs around NGC\,3268 as
tracers for the total galaxy mass. We obtained for the mass enclosed up to
$47.2\,{\rm kpc}$ from the galaxy centre a value of $2.7\times10^{12}\,{\rm M_{\odot}}$. 
This value is in reasonable agreement with the value derived from X-ray observations.
\end{itemize}

\section*{Acknowledgments}   
This work was funded with grants from Consejo Nacional de Investigaciones    
Cient\'{\i}ficas y T\'ecnicas de la Rep\'ublica Argentina, and Universidad Nacional de  
La Plata (Argentina).
TR is grateful for financial support from FONDECYT project Nr.\,1100620,  
and from the BASAL Centro de Astrof\'isica y Tecnolog\'ias Afines (CATA) PFB-06/2007. 
JP Caso is grateful to Francisco Azpillicueta, Ignacio Gargiulo and Cristian Vega Mart\'inez 
for useful discussions. 
    
\bibliographystyle{mn2e} 
\bibliography{biblio}

\label{lastpage}    
\end{document}